\documentclass[epj,nopacs]{svjour}
\usepackage{graphics}
\graphicspath{{./images/}}

\usepackage{color}
\usepackage{siunitx}
\usepackage{amsmath}
\usepackage{amssymb}
\usepackage{epstopdf}
\usepackage[colorinlistoftodos,prependcaption,textsize=tiny]{todonotes}

\begin{document}
	\title{Detecting Friedel oscillations in ultracold Fermi gases}
	\author{Keno Riechers\inst{1} \and Klaus Hueck\inst{1} \and Niclas Luick\inst{1} \and Thomas Lompe\inst{1} \and Henning Moritz\inst{1} }                     
	\mail{henning.moritz@physik.uni-hamburg.de}
	\institute{Institut f\"ur Laserphysik, Universit\"at Hamburg, Luruper Chaussee 149, 22761 Hamburg, Germany} 
	\date{Received: date / Revised version: date}
	%
	\abstract{
		Investigating Friedel oscillations in ultracold gases would complement the studies performed on solid state samples with scanning-tunneling microscopes. In atomic quantum gases interactions and external potentials can be tuned freely and the inherently slower dynamics allow to access non-equilibrium dynamics following a potential or interaction quench. Here, we examine how Friedel oscillations can be observed in current ultracold gas experiments under realistic conditions. 
		To this aim we numerically calculate the amplitude of the Friedel oscillations which a potential barrier provokes in a 1D Fermi gas and compare it to the expected atomic and photonic shot noise in a density measurement.
		We find that to detect Friedel oscillations the signal from several thousand one-dimensional systems has to be averaged. 
		However, as up to 100 parallel one-dimensional systems can be prepared in a single run with present experiments, averaging over about 100 images is sufficient.
	} 
	\maketitle
	\section{Introduction}
	\label{intro}
	Disturbing a homogeneous Fermi gas with an impurity gives rise to Friedel oscillations \cite{Friedel1958,Villain2016}. The density distribution close to the impurity shows a spatially oscillating structure which decays with increasing distance and whose periodicity is given by half the Fermi wavelength. Friedel oscillations occur e.g. in metals when the free electron gas is disturbed by the potential associated with impurity atoms. 
	They mediate long range interactions between individual impurities, which can give rise to the formation of ordered superstructures in adsorbates \cite{Tsong1973,Lau1978} and are relevant for the interactions between magnetic impurities \cite{Briner1998}. 
	Using scanning tunneling microscopy (STM) Friedel oscillations have been observed in two-dimensional and one-dimensional electron gases at surfaces of solids and have served as a tool for the measurement of bandstructures and Fermi surfaces \cite{Crommie1993,Hasegawa1993,Sprunger1997,Hofmann1997,Yokoyama1998}.
	
	While Friedel oscillations in non-interacting  systems are fully understood, the precise impact of interactions remains an open issue. Theoretical and experimental results suggest an enhancement of the oscillation amplitude for repulsive interactions \cite{Sprunger1997,Egger1995,Simion2005}, but systematic experimental studies remain to be done. Furthermore, all observations so far have reported on static Friedel oscillations since STM cannot resolve the dynamics of electronic systems. Ultracold Fermi gases \cite{Stringari2008,Ketterle2008,Zwerger2008} have the potential to contribute to both aspects of the topic: Interactions can easily be tuned via Feshbach resonances and the dynamics of these systems can be resolved due to their much longer intrinsic timescales. Yet so far, Friedel oscillations in ultracold gases \cite{Zwerger2005,Wonneberger2004,Eggert2009} have not been observed. Motivated by recent advances in the generation and study of ultracold Fermi gases we investigate the feasibility of observing Friedel oscillations in a one-dimen\-sional gas of ultracold non-interacting fermions \cite{Moritz2005,Zimmermann2011,Zwierlein2015,Kuhr2015,Greiner2015,Thywissen2015,Bloch2016,Hueck2017b}.  
	
	\section{Density distribution around an impurity potential}
	\label{sec:calculation}
	In homogeneous non-interacting fermionic systems the one-particle eigenstates of the Hamiltonian are given by plane waves characterized by a well defined momentum $\hbar k$. When inserting a localized impurity potential the plane waves are scattered, giving rise to standing wave patterns in each single-particle wavefunction. Close to an abrupt potential change the standing waves corresponding to different occupied momenta are in phase and add up to an oscillatory modification \begin{equation}
		\delta n(r) \propto \frac{\sin(2k_Fr+\eta)}{r^D},
		\label{eq:Friedeloscillations}
	\end{equation}
	of the many-body density with respect to the unperturbed density. Here, $k_F$ denotes the Fermi vector and $D$ the dimensionality of the system. The phase shift $\eta$ depends on the precise shape of the impurity potential and the dimensionality. Since the decay of Friedel oscillations is weakest in one dimension we restrict our investigations to this case. 
	
\begin{figure}
		\includegraphics{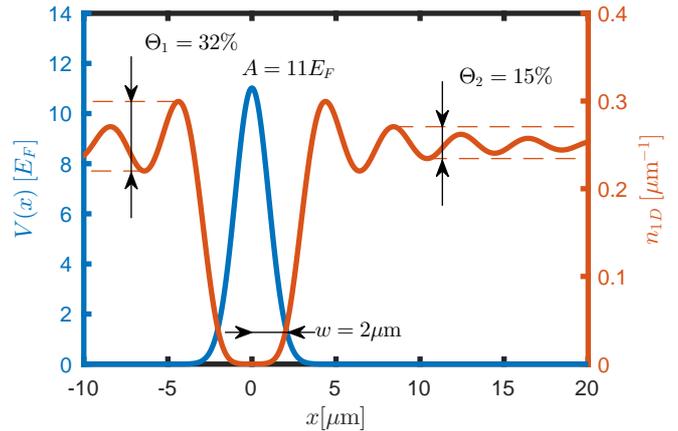}
		\caption{Density distribution of non-interacting fermions at $T=0$ around a scattering impurity in a one-dimensional and otherwise constant potential. Experimentally feasible parameters of $A=11 E_\textrm{F}$ and $w=\SI{2}{\micro \meter}$ have been chosen. Friedel oscillations with a wavelength of $\lambda_{FO}\approx\SI{4}{\micro\meter}\approx 1/\bar{n}_{1D}$ emerge symmetrically around the barrier. 
		Their peak-valley amplitudes are only slightly reduced with respect to the values $\Theta_1=\SI{37}{\percent}$ and $\Theta_2=\SI{16}{\percent}$ obtained for an infinitely sharp and high potential step.}
		\label{fig:Friedeloscillations}
	\end{figure}

	As a first step it is instructive to see how Friedel oscillations emerge for the simplest case, i.e. in a box potential of length $L$ bound by infinitely high walls for $T=0$. Filling $N$ fermions of identical spin into the lowest $N$ eigenstates yields:
	\begin{equation}
		\begin{split}
			n_\textrm{1D}(x)&=\frac{2}{L}\sum_{j=1}^{N}\sin(\pi j x/L)^2=\frac{2 N+1}{2 L}-\frac{\sin(\pi\,\frac{(2 N+1)}{L} x)}{2L\sin(\pi x/L)}\\
			&= \bar{n}_\textrm{1D}-\frac{\sin(2 k_F x)}{2L\sin(\pi x/L)}, \label{eq:1DNiFriedeloscillations}
		\end{split}
	\end{equation}
	where $\bar{n}_\textrm{1D}$ is the mean density far from the impurity and $k_F=\pi \bar{n}_\textrm{1D}$.
	This formula allows to determine the maximum amplitude of Friedel oscillations in a non-inter\-acting system. The peak-valley amplitudes of the first and second Friedel oscillation as defined in Fig. \ref{fig:Friedeloscillations} are $\Theta_1=37\%$ and $\Theta_2=16\%$. 
	
	However, due to the finite slope of the impurity potential in an experimental realization the plane waves with different momenta are reflected with different phase shifts. The standing waves are hence not in phase even close to the impurity and the amplitude of the Friedel oscillations is decreased. 
	In order to quantify this effect we perform a numerical study for a Gaussian impurity potential $V(x)$ having a height $A$ and a $1/e^2$-radius $w$:
	\begin{equation}
		V(x)=A\cdot\exp\left(-\frac{2\,x^2}{w^2}\right).
	\end{equation}
	This barrier is placed at the center of a finite size system which is limited by narrow and high potential walls outside the region of interest. We obtain the one-particle orbitals $\phi_k$ by numerically solving the discretized Schr\"odinger equation. 
	The expectation value for the particle density operator of a Fermi gas at temperature $T$ and chemical potential $\mu$ is given by \cite{Riechers2017}
	\begin{equation}
		\langle \hat{n}(x) \rangle=\sum_{k}f(\epsilon_k,T,\mu)|\phi_k(x)|^2.
		\label{eq:density}
	\end{equation}
	Here, $f$ is the Fermi distribution function and $\epsilon_k$ the energy of the orbital $\phi_k$ with wavevector $k$. 
	
	According to equation (\ref{eq:1DNiFriedeloscillations}) the wavelength of the Friedel oscillations $\lambda_{FO}=1/\bar{n}_\textrm{1D}$ is given by the inverse density and therefore equals the average particle distance. Because Friedel oscillations on scales below the resolution $R$ of the imaging system cannot be observed the maximal density in a possible experiment is constrained. Accordingly we ensure that the Friedel wavelength is 4 times larger than typical resolutions of $R=\SI{1}{\micro\meter}$ by choosing the chemical potential $\mu$ such that the average density is $\bar{n}_\textrm{1D}\approx\SI{0.25}{\micro\meter^{-1}}$.
	
	Figure \ref{fig:Friedeloscillations} shows the density distribution around a gaussian barrier at zero temperature  calculated with the approach outlined above. The peak-valley amplitudes of the first and second oscillation with respect to the average particle density are $\Theta_1=\SI{32}{\percent}$ and $\Theta_2=\SI{15}{\percent}$, respectively. 
	
	\begin{figure}
		\includegraphics{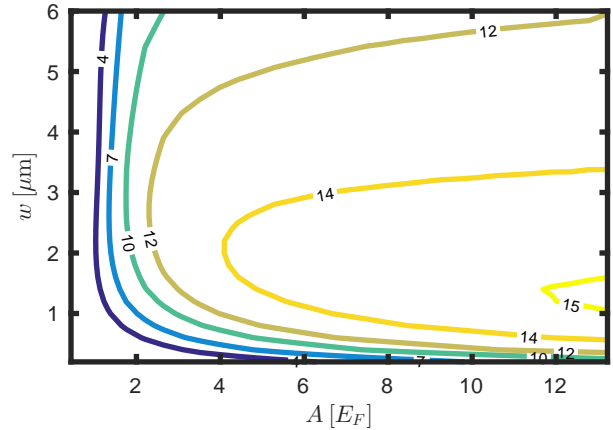}
		\caption{Dependence of the peak-valley amplitude of the 2\textsuperscript{nd} Friedel oscillation $\Theta_2$ on the parameters defining the impurity potential, namely the $1/e^2$-radius $w$ and the height $A$.  
			The color coded lines represent contour lines with equal $\Theta_2$ denoted in \%.
			For either too narrow or too low barriers no significant Friedel oscillations are present. Strongest Friedel oscillations occur for moderately narrow  ($w\simeq \SI{1.5}{\micro\meter}\approx 0.375 \lambda_{\text{FO}}$) and high ($A>12\;E_{\text{F}}$) barriers. In this regime $\Theta_2$ exceeds $\SI{15}{\percent}$.}
		\label{fig:Theta2}
	\end{figure}
	
	Since an observation of at least two density maxima is crucial for an experimental determination of $\lambda_{FO}$ we study the impact of the impurity potential height $A$ and $1/e^2$ radius $w$ on $\Theta_2$. As shown in Fig. \ref{fig:Theta2} the results range from the absence of significant Friedel oscillations for $A\lesssim E_{\text{F}}$ to $\Theta_2=\SI{15}{\percent}$ for $w=\SI{1.5}{\micro\meter}\approx0.375\;\lambda_{\text{FO}}$ and $A \gtrsim 12\;E_{\text{F}}$, where $E_{\text{F}}=h\cdot\SI{454}{\hertz}$ is defined as the Fermi energy of the unperturbed system with a density $\bar{n}_\textrm{1D}\approx\SI{0.25}{\micro\meter^{-1}}$. The amplitude of the oscillation is larger the more abrupt and pronounced the change in the potential is. Very narrow barriers allow for tunneling and therefore do not provoke strong Friedel oscillations.
	
	\begin{figure}
		\includegraphics{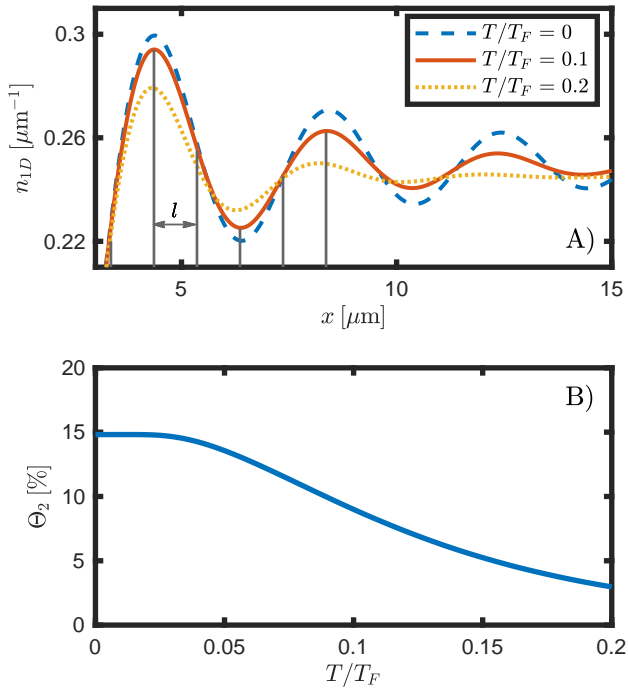}
		\caption{A) Friedel oscillations in the vicinity of the impurity potential ($w=\SI{2}{\micro\meter}$, $A=11\,E_{\text{F}}$) for different temperatures. The higher the temperature, the less pronounced are the oscillations. In the detection process, binning due to e.g. the pixel size of the camera will necessarily occur. 
		The linear size $l$ of the detection bins indicated by the grey lines must be small enough to be able to resolve the oscillation's wavelength despite the binning.
		B) The amplitude of the 2\textsuperscript{nd} oscillation $\Theta_2$ in dependence of the temperature measured in units of the Fermi temperature $T_F=E_F/k_B$. For an experimentally achievable temperature of $T/T_F=0.1$ the calculation yields $\Theta_2=\SI{9}{\percent}$.} 
		\label{fig:FiniteTemperature}
	\end{figure}
	
	In the second step of our analysis we quantitatively study the influence of finite temperature. An increase in temperature is accompanied by a loss of coherence and therefore a decrease in the amplitude of the Friedel oscillations is expected. In Fig. \ref{fig:FiniteTemperature} results on the temperature dependence are shown for the parameters $w=\SI{2}{\micro\meter}$ and $A=11E_{\text{F}}$ used also for Fig. \ref{fig:Friedeloscillations}. 
	As expected $\Theta_2$ decreases monotonously with increasing temperature. For a temperature of $T/T_F=0.1$ that can reliably achieved in quantum gas experiments $\Theta_2$ has decreased to $\SI{9}{\percent}$. Hence, the expected peak-valley amplitude of the Friedel oscillations in current experiments will most likely be limited by the finite temperature rather than the finite barrier width and height.

	\section{Experimental scenario}
	\label{sec:expsetup}
	The experimental observation of Friedel oscillations in ultracold Fermi gases requires the ability to create impurity potentials as well as to image the atomic density with a resolution on the order of half the Fermi wavelength. 
	This has become possible in recent years with the introduction of high resolution imaging using microscope objectives \cite{Zimmermann2011,Zwierlein2015,Kuhr2015,Greiner2015,Thywissen2015}. 
	Quantum gas microscopy has already enabled the study of 1D fermionic lattice systems with single atom sensitivity \cite{Bloch2016}. 
	
	In the following we describe a promising experimental scenario in which Friedel oscillations should be observable. A 2D Fermi gas of e.g. several hundred fermionic atoms (e.g.$^6$Li, $^{40}$K or $^{171}$Yb)  is prepared in a single 2D layer, which is sliced into about one hundred 1D tubes by imposing an optical lattice.
	The distance $d$ between the 1D tubes is given by the lattice spacing.
	A repulsive barrier can then be projected onto the atoms using a repulsive optical potential shaped by means of spatial light modulators (see e.g. \cite{Boyer2006,Greiner2016,Hueck2017}) providing diffraction limited feature sizes below $\SI{1}{\micro\meter}$. 
	It will remain a technical challenge to keep density deviations caused by imperfections in the potential landscape significantly smaller than the amplitude of the Friedel oscillations.

	\section{Expected signal to noise ratio}
	\label{sec:SignaltoNoise}
	As shown in Sect. \ref{sec:calculation} the observation of Friedel oscillations in ultracold Fermi gases requires the detection of signal amplitudes as small as $\SI{5}{\percent}$ of the 1D density.
	This is only possible if the signal to noise ratio (SNR) of the density measurements exceeds $20$.
	The two most important sources of noise in density images are the atomic shot noise and the detection noise. In the following we first focus on the atomic shot noise, and consider only a single one-dimensional system in order to discuss the signal to noise ratio in single atom sensitive fluorescence imaging. Finally, we calculate the signal to noise ratios achievable in absorption imaging.
		
	In order to be able to resolve the Friedel oscillations spatially, the number $\kappa$ of detection bins per Friedel wavelength should be larger than 4, otherwise the wavelength cannot be determined.
	The linear size of the bins is given by 
	\begin{equation}
		l =\frac{\lambda_{FO}}{\kappa}=\frac{1}{\kappa n_\textrm{1D}}.
		\label{eq:xi}
	\end{equation}
	
	As the average interatomic distance equals the wavelength of the Friedel oscillation, the average number of atoms located within the area of a single bin is $\langle N\rangle=1/\kappa$. 
	The atomic shot noise  $\sigma_N$ is approximately $\sigma_N=\sqrt{\langle N\rangle}$ \cite{NoiseSuppression},  yielding a relative atomic shot noise per bin of $\sigma_N/\langle N\rangle=\sqrt{\kappa}$ which is independent of the density. 
	Even for a minimal number of bins of $\kappa=4$ and no further noise sources the relative noise would be $\SI{200}{\percent}$. 
	This shows that suppressing the atomic shot noise to a relative level of $\SI{5}{\percent}$ requires an average over 1600 measurements from individual 1D systems. 
	For state of the art fluorescence imaging no further significant detection noise is added.
	Here, a very deep optical lattice is used to pin the atoms to one site during detection and single atom, single site sensitive detection is achieved \cite{Zwierlein2015,Kuhr2015,Greiner2015,Thywissen2015,Bloch2016}.
	We note that for this detection method the mean interparticle distance along the tubes must be at least $\kappa$ times larger than the optical pinning lattice spacing of typically \SI{0.5}{\mu\meter} in order to be able to spatially resolve the Friedel oscillations.
	Since in the proposed experimental setup up to 100 parallel tubes can be prepared and imaged in each realization, only data from 10 to 100 separate runs would have to be averaged. Density measurements with such sensitivity have already been performed by averaging over 1000 tubes in a bosonic quantum gas microscope setup \cite{Bloch2012}.

	Most quantum gas experiments rely on measuring two-dimensional column densities via absorption imaging. Here, the detection noise becomes relevant and is mainly caused by photon shot noise. 
	In the following we perform a calculation of the signal to noise ratio including photon shot noise. We find that the major limitation is still given by the atomic shot noise and that photon shot noise reduces the signal to noise ratio by at most a factor of 1.25.
	
	In absorption imaging the 2D density $n_\textrm{2D} = n_\textrm{1D}/d$ is measured rather than the 1D density. For the calculation we choose the detection bins to be two-dimensional pixels with pixel lengths $l$ along the tube direction and $d$ perpendicular to it, where $d$ is the distance between tubes.
	This ensures that effectively only one 1D system is measured per row of pixels, despite the fact that the tube structure proposed in Sect. \ref{sec:expsetup} cannot be resolved for typical lattice spacings $d\approx\SI{0.5}{\micro\meter}$ between the tubes. 
	Enlarging the pixel size perpendicular to the tube direction would be analogous to averaging over several parallel 1D systems. 
	The average number of atoms per pixel is $\langle N_\textrm{pix}\rangle=n_\textrm{2D}l  d=\kappa^{-1}\leq1/4$ and the atomic shot noise approximately $\sigma_{N_\textrm{pix}}=\kappa^{-1/2}$ yielding $\sigma_{n_\textrm{2D}}=\kappa^{-1/2}/l d$ per pixel. 
	
	In absorption imaging the 2D density is measured indirectly by determining the number of photons $p_\textrm{sc}$ scattered by the atoms 
	\begin{equation}
		p_\textrm{sc}=p_\textrm{ref}-p_\textrm{a}=p_\textrm{ref}-p_\textrm{in}T.
		\label{eq:ScatteredPhotons}
	\end{equation}
	Here $p_\textrm{a}$  denotes the number of photons transmitted by the atoms when illuminated by $p_\text{in}$ photons and $p_\textrm{ref}$ is the number transmitted in the absence of atoms. $p_\text{in}$ and $p_\text{ref}$ originate from identically prepared laser pulses and have the same mean value, but are stochastically independent with $\sigma_{p_{\textrm{in},\textrm{ref}}}=\sqrt{p_\textrm{in}}$.
	The transmission coefficient $T=\exp(-\beta n_\textrm{2D})$ can be a approximated as $T\simeq 1-\beta n_\textrm{2D}$ in the limit of low optical densities $\beta n_\textrm{2D}\ll 1$ which is relevant here. $\beta$ denotes the scattering cross section of the corresponding atomic transition.
	
	Eq. \ref{eq:ScatteredPhotons} shows that it is convenient to regard the number of scattered photons $p_\textrm{sc}\approx p_\textrm{in}\beta n_\textrm{2D}$ as the relevant signal and to compare it with the corresponding standard deviation $\sigma_\textrm{sc}$ to determine the signal to noise ratio of the density measurement $ \textrm{SNR}=p_\textrm{sc}/\sigma_\textrm{sc}$.
	Gaussian error propagation yields
	\begin{equation}
		\begin{split}
			\sigma^2_\textrm{sc}&=\sigma_{p_{\textrm{ref}}}^2+(1-\beta n_\textrm{2D})^2\sigma_{p_\textrm{in}}^2+p_\textrm{in}^2\beta^2\sigma_{n_\textrm{2D}}^2\\
			&=p_\textrm{in}+(1-\beta n_\textrm{2D})^2p_\textrm{in}+p_\textrm{in}^2\beta^2\sigma_{n_\textrm{2D}}^2 
		\end{split}
	\end{equation}
	for the variance of the scattered photons. The signal to noise ratio then reads
	\begin{equation}
		\textrm{SNR}\approx \frac{p_\textrm{in}\beta n_\textrm{2D}}{\sqrt{(2-2\beta n_\textrm{2D})p_\textrm{in}+p_\textrm{in}^2\beta^2\sigma_{n_\textrm{2D}}^2}}.
		\label{eq:SNR}
	\end{equation}
	
	For very high numbers of incoming photons - i.e. in the limit of vanishing relative photon shot noise - the SNR is ultimately limited by
	\begin{equation}
		\lim_{p_\textrm{in}\rightarrow\infty} \textrm{SNR}=\frac{n_\textrm{2D}}{\sigma_{n_\textrm{2D}}} =\frac{N_\textrm{pix}}{\sigma_{N_\textrm{pix}}}=\frac{1}{\sqrt{\kappa}}
	\end{equation}
	and hence limited by atomic shot noise as in the case of single atom sensitive fluorescence imaging.
	
	\begin{figure}
		\includegraphics{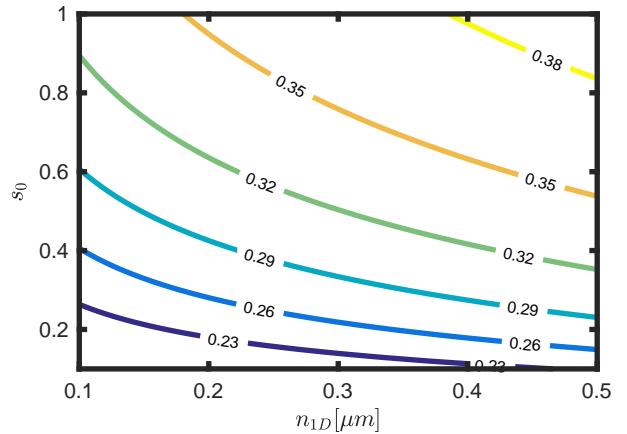}
		\caption{Signal to noise ratios achievable for a density measurement on a single pixel with absorption imaging as a function of the one-dimensional density $n_\textrm{1D}$ and the imaging saturation $s_0$ for $^6$Li and a lattice spacing of $d=\SI{0.5}{\micro\meter}$. The imaging time has been adapted to optimize the SNR for each $s$ and $n_\textrm{1D}$ and the pixel length chosen such that a Friedel oscillation extends over 4 pixels. In the displayed region the SNR increases for both, increasing density and saturation. The saturation was limited to $s_0<1$ since the approximations made in the calculation fail for higher saturation.}
		\label{fig:SNR_Li}
	\end{figure}
	
	It is therefore desirable to work with high intensities $I$ and long illumination times $t$. 
	However, the number of photons that can be scattered by an individual atom is limited by motional blurring. When scattering photons for an extended time $t$ at an intensity $I$ an atom performs a random walk in momentum space. This leads to a motional blurring $\delta x(t,I)$ of its position with respect to its original position.
	To ensure that the density distribution is not altered significantly during the imaging process the condition $\delta x(t,I)\leq l$ should be fulfilled. 
	The full calcuation \cite{Horikoshi2016} yields an upper bound $t_\textrm{max}$ for the illumination time
	\begin{equation}
		t\leq 3\left(\frac{l\,\lambda \,m}{2 h}\sqrt{\frac{1+s_0}{\Gamma s_0}}\right)^{2/3}\equiv t_\textrm{max}.
		\label{eq:tau}
	\end{equation}
	Here the saturation parameter $s_0=I/I_{sat}$ is used where $I_{sat}$ refers to the saturation intensity and $m$ to the atomic mass. $\Gamma$, $\lambda$ and $\nu$ are the linewidth, wavelength and frequency of the atomic transition. The optimal signal to noise ratio is achieved for the maximal illumination time $t_\textrm{max}$ and its dependence on $n_\textrm{2D},s_0$ and $\kappa$ can be calculated by using Eq. (\ref{eq:SNR}) and 
	\begin{equation}
		p_\textrm{in}=\frac{s_0 I_{sat}}{h\nu}l\,  d\, t_{\text{max}}.
	\end{equation}
	We evaluate the optimal signal to noise ratio for an experimentally accessible configuration, i.e. for $^6\text{Li}$ atoms, a lattice spacing of $d=\SI{0.5}{\micro\meter}$ and $\kappa=4$. The results for varying density and saturation are presented in Fig. \ref{fig:SNR_Li}.
	For any saturation the signal to noise ratio improves with increasing density but care must be taken that the 1D density is such that the Friedel wavelength remains $\kappa$ times larger than the optical resolution. 
	Within the parameter range considered here a maximal SNR per pixel of $\textrm{SNR}\approx0.39$ can be achieved for a light atom such as $^6$Li. For heavier atoms such as $^{40}$K the signal to noise ratio approaches the atomic shot noise limit of $\textrm{SNR}=1/\sqrt{\kappa}= 0.5$. 
	
	\section{Conclusion}
	In this article we study the feasibility of observing Friedel oscillations in ultracold one-dimensional Fermi gases. 
	We numerically calculate the amplitude of the density oscillations for a suitable experimental setup and find that for currently achievable temperatures of $T/T_F=0.1$ it is on the order of $\SI{10}{\percent}$ of the total density.
	We then calculate the expected noise for a density measurement on a single 1D system, which is limited by atomic shot noise and exceeds the amplitude of the Friedel oscillations by a factor of 20.  
	Nevertheless, since many 1D systems can be observed in a single run the noise amplitude can be sufficiently reduced by averaging over 100 images.
	Therefore we conclude that an observation of Friedel oscillations is experimentally feasible.
	This would open up the possibility to investigate their non-equilibrium dynamics and to use them to probe Fermi liquids with attractive and repulsive interactions.
	
	\section{Authors contributions}
	All authors  were involved in the discussion of the physical setup and the interpretation of the numerical results.
	The calculations were mainly performed by K.R..
	All the authors were involved in the preparation of the manuscript.
	
	\acknowledgement{The research leading to these results has received funding from the European Union’s Seventh Framework Programme (FP7/ 2007-2013) under grant agreement No. 335431 and by the DFG in the framework of SFB 925 and the excellence cluster the Hamburg Centre for Ultrafast Imaging CUI. We thank T. Giamarchi, L. Mathey and F. Werner for stimulating discussions.}
	
	%

\begin{thebibliography}{}
		%
		%
		\bibitem{Friedel1958}
		J. Friedel, Del Nuovo Cimento \textbf{2}, 287 (1958)
		\bibitem{Villain2016}
		J. Villain, M. Lavagna, P. Bruno, Comptes Rendus Physique \textbf{17}, 302 (2016)
		
		\bibitem{Zwerger2005}
		A. Recati, J. N. Fuchs, C. S. Peca, W. Zwerger. Phys. Rev. A \textbf{72}, 023616 (2005)
		
		\bibitem{Tsong1973}
		T. T. Tsong, Phys. Rev. Lett. \textbf{31}, 1207 (1973)
		\bibitem{Lau1978}
		K. H. Lau, W. Kohn, Surface Science \textbf{75}, 69 (1978)
		
		\bibitem{Briner1998} B. Briner, P. Hofmann, M. Doering, H. P. Rust, E. Plummer, A. Bradshaw, Phys. Rev. B \textbf{58}, 13931 (1998)
		
		\bibitem{Crommie1993}
		M. F. Crommie, C. P. Lutz, D. M. Eigler, Nature \textbf{363}, 524 (1993)
		\bibitem{Hasegawa1993}
		Y. Hasegawa, P. Avouris, Phys. Rev. Lett. \textbf{71}, 1071 (1993)
		\bibitem{Sprunger1997}
		P. T. Sprunger, L. Petersen, E. W. Plummer, E. Laegsgaard, F. Besenbacher, Science \textbf{275}, 1764 (1997)
		\bibitem{Hofmann1997}
		P. Hofmann, B. Briner, M. Doering, H. P. Rust, E. Plummer, A. Bradshaw, Phys. Rev. Lett. \textbf{79}, 265 (1997)
		\bibitem{Yokoyama1998}
		T. Yokoyama, M. Okamoto, K. Takayanagi, Phys. Rev. Lett. \textbf{80}, 3423 (1998)
		\bibitem{Egger1995}
		R. Egger, H. Grabert, Phys. Rev. Lett. \textbf{75}, 3505 (1995)
		\bibitem{Simion2005}
		G. E. Simion, G. F. Giuliani, Phys. Rev. B \textbf{72}, 045127 (2005)
		
		
		\bibitem{Stringari2008}
		S. Giorgini, L. P. Pitaevski, S. Stringari, Rev. Mod. Phys. \textbf{80}, 1215 (2008)
		\bibitem{Ketterle2008}
		W. Ketterle, M. W. Zwierlein, in \textit{Proceedings of the International School of Physics: Enrico Fermi, Course CLXIV, Varenna 2006}, edited by M. Inguscio, W. Ketterle, C. Salomon (IOS Press, Amsterdam, 2008), p. 95
		\bibitem{Zwerger2008}
		I. Bloch, J. Dalibard, W. Zwerger, Rev. Mod. Phys. \textbf{80}, 885 (2008)
		
		\bibitem{Wonneberger2004}
			S. N. Artemenko, G. Xianlong, W. Wonneberger, J. Phys. B \textbf{37}, S49 (2004) 
		\bibitem{Eggert2009}
		S. A. S\"offing, M. Bortz, I. Schneider, A. Struck, M. Fleischhauer, S. Eggert, Phys. Rev. B \textbf{79}, 195114 (2009)
		
		\bibitem{Moritz2005}
		H. Moritz, T. St\"oferle, K. G\"unter, M. K\"ohl, T. Esslinger, Phys. Rev. Lett. \textbf{94}, 210401 (2005)
		\bibitem{Zimmermann2011}
		B. Zimmermann, T. Mueller, J. Meineke, T. Esslinger, H. Moritz, New J. Phys. \textbf{13}, 043007 (2011)
		\bibitem{Zwierlein2015}
		L. W. Cheuk, M. A. Nichols, M. Okan, T. Gersdorf, V. V. Ramasesh, W. S. Bakr, T. Lompe, M. W. Zwierlein, Phys. Rev. Lett. \textbf{114}, 193001 (2015)
		
		\bibitem{Kuhr2015}
		E. Haller, J. Hudson, A. Kelly, D. A. Cotta, B. Peaudecerf, G. D. Bruce, S. Kuhr, Nature Phys. \textbf{11}, 738 (2015)
		
		\bibitem{Greiner2015}
		M. F. Parsons, F. Huber, A. Mazurenko, C. S. Chiu, W. Setiawan, K. Wooley-Brown, S. Blatt, M. Greiner, Phys. Rev. Lett. \textbf{114}, 213002 (2015)
		
		\bibitem{Thywissen2015}
		G. J. A. Edge, R. Anderson, D. Jervis, D. C. McKay, R. Day, S. Trotzky, J. H. Thywissen, Phys. Rev. A \textbf{92}, 063406 (2015)
		
		\bibitem{Bloch2016}
		M. Boll, T. A. Hilker, G. Salomon, A. Omran, J. Nespolo, L. Pollet, I. Bloch, C. Gross, Science \textbf{353}, 1257 (2016)
		
		\bibitem{Hueck2017b}
		K. Hueck, N. Luick, L. Sobirey, J. Siegl, T. Lompe, H. Moritz, manuscript in preparation (2017)
		
		
		\bibitem{Riechers2017}
		K. Riechers, M. Sc. thesis, University of Hamburg (2017)
		
		\bibitem{Boyer2006}
		V. Boyer, R. M. Godun, G. Smirne, D. Cassettari, C. M. Chandrashekar, A. B. Deb, Z. J. Laczik, C. J. Foot,
		Phys. Rev. A \textbf{73}, 031402 (2006)
		
		\bibitem{Greiner2016}
		P. Zupancic, P. M. Preiss, R. Ma, A. Lukin, M. E. Tai, M. Rispoli, R. Islam, M. Greiner, Opt. Exp. \textbf{24}, 13881 (2016)
		
		\bibitem{Hueck2017}
		K. Hueck, A. Mazurenko, N. Luick, T. Lompe, H. Moritz, Rev. Sci. Instrum. \textbf{88}, 016103 (2017)
		
		\bibitem{NoiseSuppression}
		The suppression of atomic shot noise for low temperature Fermi gases due to antibunching \cite{Mueller2010} is only significant for detection volumes larger than the correlation length, which is not the case here.
		
		\bibitem{Mueller2010}
		T. M\"uller, B. Zimmermann, J. Meineke, J.-P. Brantut, T. Esslinger, H. Moritz, Phys. Rev. Lett. \textbf{105}, 040401 (2010)
		
		\bibitem{Bloch2012}
		M. Cheneau, P. Barmettler, D. Poletti, M. Endres, P. Schau\ss{}, T. Fukuhara, C. Gross,
		I. Bloch, C. Kollath, S. Kuhr, Nature \textbf{481}, 484 (2012) 
		
		\bibitem{Horikoshi2016}
		M. Horikoshi, A. Ito, T. Ikemachi, Y. Aratake, M. Kuwata-Gonokami, M. Koashi, arXiv:1608.07152 (2016)
		
	\end{thebibliography}
	%

\end{document}